Numerical evaluation of Chandrasekhar's H-function , its first and second differential coefficients , its pole and moments from the new form for plane parallel scattering atmosphere in Radiative transfer . ***

1.*Rabindra Nath Das ,Visiting Faculty Member ,
*Email address : rndas1946@yahoo.com
2.**Dr. Rasajit Bera, Head of the Department ,
**Email address : rasajit@yahoo.com

Department of Mathematics , Heritage Institute of Technology , Chowbaga Road, Anandapur , P.O.East Kolkata Township, Kolkata-7000107, West Bengal , India.

**Abstract:**

In this paper, the new forms obtained for Chandrasekhar's H- function in Radiative Transfer by one of the authors both for non-conservative and conservative cases for isotropic scattering in a semi-infinite plane parallel atmosphere are used to obtain exclusively new forms for the first and second derivatives of H-function . The numerics for evaluation of zero of dispersion function , for evaluation of H-function and its derivatives and its zeroth , the first and second moments are outlined. Those are used to get ready and accurate extensive tables of H-function and its derivatives , pole and moments for different albedo for scattering by iteration and Simpson's one third rule . The schemes for interpolation of H-function for any arbitrary value of the direction parameter for a given albedo are also outlined. Good agreement has been observed in checks with the available results within one unit of ninth decimal.

**1.Introduction:**

The basic equation of radiative transfer or neutron diffusion or diffusion of electron in metallic lattice is a linear integro differential equation . In semi-infinite plane parallel scattering atmosphere its solution gives the emergent intensity from the bounding face in terms of H-function of Chandrasekhar[1] .This H-function plays an important role to understand the Schuster model. on scattering and the Eddington - Milne model (cf..Chandrasekhar[1]). of scattering and absorption arising from ionization and recombination and to interpret the absorption lines in stellar spectra and to determine the contours of the residual intensity in the line and the thickness of the reflector for building the nuclear reactor

The tabular values of H-function and tables of its moments and differential coefficients are also used by Hapke [2],[3],[4] to model the reflection of light by particulate surface material of celestial bodies such as planets , moons, and asteroids . Those are also used by Van de Hulst [5],[6] to study the radiation transport in spherical clouds and in his theory of light scattering for optically large spheres .

Those are also used to interpret the model of Maxwell on basic slip problem in kinetic theory for the rarefied gas flows and also to understand the Kramers' problem (velocity slip problems). In kinetic theory , the H function has a direct use in problems of condensation and evaporation of molecules from surfaces , drops and particles (cf.,Loyalka and Naz [7]) .

The H-functions have also practical applications to understand better the transport of radiation in tissue in connection with the optical tomography in infrared region , dispersal of laser beam in a medium and the emissive power of an industrial surface .

The detailed numerical application of the mathematical theory on radiative transfer depends , however upon the availability of the H- function in sufficient tabular level. This requirement has been initially met by the heroic computation of Chandrasekhar and Breen [9]for isotropic and anisotropic scattering using the nonlinear equation for H-functions. This happens to be the first tabular values of H-functions for different albedo for single scattering and for different direction parameters both for conservative and non conservative cases .

Placzek and Seidel [10] and Placzek[11] have computed the H- functions from the Wiener-Hopf integral of H- functions for conservative cases only and given the table of H- function for different direction parameters .

Stibbs and Weir [12] have obtained H-function by the direct quadrature of an integral for both non conservative and conservative cases and described numerical procedures which have enabled the accuracy of the calculations to be maintained in the neighborhood of singularities in the integrand and its first derivative. Those are happened to be the second acceptable tables of H-function and its first moment for different values of albedo for scattering and for different direction parameters .

  Kourganoff [13] , Busbridge[14a, 14b] have obtained some new forms of H-functions from the dispersion function T(z) and linear integral equation of H-function using Wiener- Hopf technique in complex z plane and obtained some numerical results to testify their analytical forms and compared those test results with those of Chandrasekhar[1] , Placzek and Seidel[10] , Placzek[11] , Stibbs and Weir [12].

  Fox [14c]has reduced the non –linear integral equation to linear integral equation for having solution as a Riemann Hilbert problem and obtained a form which is not amenable for numerical computation.

  Zweifel [15], Shulties and Hill[16] , Siewert [17], Garcia and Siewert [18], Barichello and Siewert [19a,19b] also have obtained the form of H(z) using the Case 's [20] eigen function approach and obtained a new form of H-function in terms of a new analytic function X(z) .

Zelazny [21] , Sobolev [22,23]), Karanjai and Sen [24], Ivanov[25] , Domke[26] have also worked on H- function to give approximate forms for evaluation.

Dasgupta[27] has used Wiener-Hopf technique to the dispersion function T(z) and obtained an explicit form of H(z) from the dispersion function T(z) and the linear integral equation .He has outlined numeric for numerical evaluation of H(z) and approximated the kernel of the integral by Legendre polynomials and compared the numerical values of H(z) with those of Stibbs and Wier[12] both for non conservative and conservative cases .

Dasgupta [28] has obtained a new representation of Chandrasekhar's H- function using Wiener Hopf technique to dispersion function T(z) and the linear integral equations.He has obtained H(z) separating poles and branch points as sum of two functions .

Das[29]has used the Laplace transform to the basic integro differential equation and obtained linear integral equation for emergent intensity from the bounding face in terms of H-function and obtained the same analytical form of H-function of Dasgupta[28]by inversion of Laplace Transform without any scheme for numerical computation of H(z) .

Islam and Dasgupta [30]have considered the H- function of Dasgupta[28] as an eigen value problem and they represented the symmetric kernel of the Fredholm integral equation of second kind into degenerate kernel through finite Taylor's expansion in terms of eigen values and eigen functions and evaluated the H- function numerically and compared the numerical values in test cases with the numerical solution obtained so far .

.Hovenier, Mee and Heer [31]have used the non linear equation of H-functions to determine its tables and obtained from the non linear form of H(z) the forms of first derivative H$'$(z) and second derivative H$''$(z) and computed the tables of those by Gausian interpolation and compared the numerical results with those of Chandrasekhar[1] and Placzek [11], Stibbs and Weir[12] for H- function .They are the first to obtain the tables of $H^{/}(z)$ and $H^{//}(z)$ and their graphs .

Bergeat and Rutily [32a,32b], Rutily [33] has obtained a new form of H-function by using Cauchy integral equations and some auxiliary functions .

Wegert [34] has also outlined a method for solution of the linear equation for H-function.

Das[35] has used the theory of linear singular integral equations and Wiener-Hopf technique to the linear integral equation of H(u) and Das[36] has also used the theory of linear singular integral equation and the theory of Riemann - Hilbert problem and obtained in both methods the same simplest form of H(u) on separating the pole and branch points of H(u ) .Some numerics have also presented there for evaluation of H(u), and its moments both for non-conservative and conservative cases.

In this paper , the new forms obtained for H(u) function of Chandrasekhar in radiative transfer by one of the author in Das [35,36] both for non-conservative and conservative cases for isotropic scattering in a semi-infinite plane parallel atmosphere are used to obtain new forms for $H^{/}(u)$ and $H^{//}(u)$ .

Restricting ourselves to isotropic scattering , the main purposes of this paper are : i) to provide easy numerics for computation of zero of dispersion function by simple iteration ; ii) to establish the numerics for numerical evaluation of this new form of H- functions by Simpson's one third rule ;iii) to provide numerical tables of H(u) for different values of particle albedo with different direction parameters ;iv) to present the new forms of $H^{/}(u)$ and $H^{//}(u)$ which are amenable for easy evaluation by Simpson's one third rule ; v) to present numerics for numerical evaluation of $H^{/}(u)$ and $H^{//}(u)$ for different values of particle albedo; vi) to provide numerical tables for $H^{/}(u)$ , $H^{//}(u)$ for different values of albedo for single scattering using Simpson's one third rule ; vii) to present the tables of zeroth , first and second moments, extrapolation distance (ratios of the second moment to the first moment of H(u)) for different values of albedo ; viii) to provide schemes of interpolation of H(u) for any arbitrary u for a particular albedo using the tables of the first and second derivatives of H(u) ; ix) to compare the numerical results of H(u) , $H^{/}(u)$ and $H^{//}(u)$ with the results available for H(u),$H^{/}(u)$ and $H^{//}(u)$ . Good agreement has been

observed in checks within one unit of ninth decimal with that computed by

Chandrasekha and Breen[9], Placzek [11] and Stibbs and Weir[12],Ivanov [25]

and Hovenier,Mee and Hee [31] .

## 2.Mathematical Background:

The basic equation of radiative transfer in plane parallel scattering atmosphere is a linear integro differential equation . In semi-infinite atmosphere, the emergent intensity from the bounding face in the direction $\cos^{-1} u$ is obtained in terms of Chandrasekhar's H - function where $H(u)$ satisfies the non linear integral equation as follows:

$$H(u) = 1 + u\,H(u) \int_0^1 U(t)\,H(t)\,d\,t\,/\,(t+u) \;\; ; \;\; 0 \le u \le 1 \qquad (1)$$

where , in astrophysical contexts ,

$U(u)$ is assumed to be an even , real , nonnegative bounded function in the interval $0 \le u \le 1$ and satisfies the condition

$$U_0 = \int_0^1 U(u)\,d\,u \; \le 1/2 . \qquad (2)$$

The case of equality in (2), i.e. $U_0 = \frac{1}{2}$ is called the conservative case, i.e., only when there is no true absorption of radiation and efficiency of scattering is unity and inequality, i.e. , $U_0 < \frac{1}{2}$ is called the non-conservative case, i.e., on each scattering , more radiation was not emitted than was incident.

An alternative non linear form of the integral equation (1) of $H(u)$ which is more suitable for numerical evaluation has been obtained by Chandrasekhar [1] as

$$1/H(u) = (1 - 2\,U_0)^{1/2} + \int_0^1 t\,U(t)\,H(t)\,d\,t\,/\,(t+u) \;\; , \;\; 0 \le u \le 1 \qquad (3)$$

Chandrasekhar [1] has found that the dispersion function ,$T(z)$ from the non linear integral equation (1) as

$$T(z) = 1 - 2\,z^2 \int_0^1 U(t)\;d\,t\,/\;(z^2 - t^2) \;\; ; \qquad (4)$$

Here $T(z)$ is an even function defined in the complex z plane cut along $(-1,1)$ with branch points at $z=-1$ and $z=1$ and has only two zeros at infinity when $U_0 = \frac{1}{2}$ and has only two real zeros at $\mathrm{Re}(z) = +1/k$ and $\mathrm{Re}(z) = -1/k$ where $0 < k < 1$ in the z plane cut along $(-1,1)$ when $U_0 < \frac{1}{2}$ .
If $H(z)$ is a solution of equation (1) which is continuous in the interval $0 \le \mathrm{Re}(z) = u \le 1$ , regular in the z plane cut along $(-1,0)$ then $1/H(z)$ is also continuous in the interval $0 \le \mathrm{Re}(z) = u \le 1$ , regular in the z plane cut along $(-1,0)$ . In the z plane cut along $(-1,1)$ , $H(z)$ satisfies

$$H(z)\,H(-z) = (T(z))^{-1} \;\; ; \qquad (5)$$

and it may be determined from an integral of the form :

$$\log H(u) = (2\pi i)^{-1} \int_{-i\infty}^{i\infty} \log(T(w)) z \, dw / (w^2 - u^2) \quad ; \quad (6)$$

Chandrasekhar and Breen[9] have observed that the representation of H(u) as a definite integral (cf. equations (6) ) may be used to evaluate H- function but in practice they have tried to solve the non-linear integral equations (1) and (3) of the H- functions directly by a process of iteration and they have observed that the non linear integral equation (3) is more suitable for the purpose of their iteration than that of the equation (1).They have obtained tables of H- functions for various values of $U(u)= c/2$ , $0 \le c \le 1$ using equation (3) . They have opined that it is suitable to start iteration with third approximation for H(u) in terms of the Gaussian interpolation and characteristic roots of T(z) in non conservative cases though in conservative cases it is preferable to start with the fourth approximation .Those are happened to be the first tables of H-functions .

The problem of the determination of the angular distribution of the emerging neutrons has been solved by Wiener and Hopf (cf. Kourganoff [13] ) and they have found the emergent distribution function $\Phi(u)$ as Wiener-Hopf integral in conservative cases as

$$\Phi(u) = \tfrac{1}{2} (1+u)$$

$$x \exp\left\{ u\pi^{-1} \int_0^{\pi/2} \log[\sin^2 x / (1 - x \cot x)] \, dx / (\cos^2 x + u^2 \sin^2 x) \right\} . \quad (7)$$

Placzek [11] has used the form of Wiener – Hopf integral (cf. e q. (7) ) and derived the simplified expression of H(u) in conservative case only (i.e. with U(u) =1/2 and $U_0 = \tfrac{1}{2}$ ) and evaluated the H(u) numerically with

$$H(u) = (1+u)$$
$$x \exp\left( u\pi^{-1} \int_0^{\pi/2} \log(\sin^2 x / (1 - x \cot x)) \, dx / (1-(1-u^2)\sin^2 x) \right); \quad (8)$$

and prepared a table of H( u) for conservative cases of isotropic scattering .

Placzek and Seidel [10] have used the Laplace Transform and Wiener Hopf technique to find an asymptotic solution of Milne problem(cf.. Chandrasekhar [1] )in transport theory and derived the simplified expression for extrapolated distance $z_9$ ( the distance where the emergent intensity of radiation beyond the barrier medium reduces to zero ) .as

$$z_0 = 6/\pi^2 + \pi^{-1} \int_0^{\pi/2} (3/x^2 - 1/(1 - x \cot x)) \, dx , \quad (9)$$

$$= 0.71044609. \quad (10)$$

Stibbs and Weir [12] have used integral form of H(u) function ( cf. equation (6) ) for direct quadrature of the integral both for non conservative and conservative cases for U(u) = c/2 , 0<c ≤1 where c is called the albedo for isotropic scattering . The form is :

$H(u,c) = \exp( I(u ,c))$ ,

where

$$I(u,c) = -u \pi^{-1} \int_0^{\pi/2} \log ( 1 - c x \cot x ) \, dx / (\cos^2 x + u^2 \sin^2 x ) ; \quad (11)$$

They have obtained the accuracy of the calculations considering the neighborhood of singularities in the integrand and its first derivative. Those have happened to be the second acceptable tables of H(u) for non conservative and conservative cases and also for the first moment of the H-function only.

C.Fox [14c] has reduced the non linear integral equation (1) to a linear integral equation of the form :

$$T(z) \, H(z) = 1 + z \int_0^1 U(x) H(x) \, dx / ( x-z ) , \quad (12)$$

where z is complex in the plane cut along (-1,1) .

and then to linear singular integral equation of the form :

$$T_o (x) H (x) = 1 + x \, P \int_0^1 U(u) H(u) \, du / ( u - x ), \; 0< x< 1; \quad (13)$$

where $T_0(x) = 1 - 2 x^2 \int_0^1 ( U(u) - U(x) ) \, du / (x^2 - u^2 )$

$$- x \, U(x) \log (( 1+x) /(1-x) ) ), \quad (14)$$

where P is interpreted to be a Cauchy' principal value of the integral and obtained an analytical expression for H(x) considering it as a Riemann Hilbert Problem

Busbridge [14a,14b] has used the linear integral equation (12) of H(z) and she has found that all the solutions of equation (12) will not satisfy equation (1) in cases $U_0 = \frac{1}{2}$ and $U_0 < \frac{1}{2}$ and finally she has found a family of solutions of equation (12) .

Das [35] has derived a new expression for H(z) as solution of linear integral equation (12) with its existence and uniqueness using Wiener- Hopf technique and theory of linear singular integral equations as

### i) for non conservative cases ($U_0 < 1/2$) :

$$H(z) = A^{-1/2}(1+z) \exp\left(-\int_0^1 \theta(t)\, d\,t \,/\,(t+z)\right)\,/\,(1+k\,z)\,, \quad (15a)$$

$$\theta(t) = \pi^{-1}\tan^{-1}(\pi t\, U(t)\,/\,T_0(t))\,, \quad\quad\quad\quad (15b)$$

$$A = D/k^2\,; \quad\quad\quad\quad\quad\quad (15c)$$

$$D = 1 - 2\,U_0\,; \quad\quad\quad\quad\quad\quad (15d)$$

$k$ , ($0 \leq k < 1$ ) is the zero of the dispersion function $T(z)$ (e.g equation(4))

and

### ii) for conservative case( $U_0 = 1/2$ ) :

$$H(z) = C^{-1/2}(1+z) \exp\left(-\int_0^1 \theta(t)\, d\,t \,/\,(t+z)\right)\,, \quad\quad (16a)$$

$$\theta(t) = (\pi)^{-1}\tan^{-1}(\pi t\, U(t)\,/\,T_0(t))\,; \quad\quad (16b)$$

$$C = 2\,U_2\,; \quad\quad\quad\quad\quad\quad (16c)$$

$$U_2 = \int_0^1 x^2 U(x)\, d\,x\,, \quad\quad\quad\quad (16d)$$

where
$T_0(t)$ is given by equation (14 ) and also has outlined numerics for evaluation of H – function and its integral properties.

Das[36] has also found the same forms of H(z) (e.g equations 15a, 16a ) from equation (12) and (13) both for non conservative and conservative cases of isotropic scattering considering it as a solution of Riemann Hilbert Problem using the theory of linear singular integral equations and contour integration and justified the equivalence of this method with that of Wiener-Hopf technique in Das [ 35] in derivation of solution of equations (12) and (13) .

### 3.Application aspect of H-function   :

#### A) Moments of H-function , extrapolatin distanceand numerics therefor :

 In his pioneering paper on interpretation of absorption lines  in stellar spectra , Schuster( eg Chandrasekhar[1] has introduced  an idealization of a stellar atmosphere which has proved extremely useful in preliminary investigation of many astrophysical problems. A quantity of particular interest in the application of Schuster model to the interpretation of the contours of the absorption lines in stellar spectra . This is the ratio of the emergent flux in the lines  to the outward flux  which represent the background continuum . This ratio defines the residual intensity in the line. In Schuster model , the contours of absorption lines for semi infinite atmosphere  and the residual intensity in the line are  dependent on both the first moment and second moment of the H-function.

 In Schuster model ,  the continuous absorption coefficient  is never negligible  compared to the line scattering coefficient . A correct theory of stellar absorption lines must therefore include both the effect of line scattering and continuous absorption .  Eddington and Milne have taken into consideration of the  effects of scattering   and absorption in their model  .In the Eddington  - Milne model ( cf. Chandrasekhar [1] ) for the interpretation of the contours of the scattering and absorption lines in semi-infinite atmosphere and to the determination of residual intensity of  the emergent flux in the line ,the residual intensity in the line  is dependent on the first moment of the H-function but the residual intensity in the emergent flux  is dependent on the both the first and second moment of the  H- function . Hence the moments of the H-functions are very important in determination of contours of the scattering and absorption lines in the spectrum and also for determination of the residual intensity of the emergent flux in the lines  .

 Maxwell  has constructed first approximate solutions of basic slip problems in the rarefied gas flows by distinguishing between the incident and out going distribution at the surface For the velocity slip problem ( also known now as the Kramers' problem ), he has distinguished between the incident and outgoing molecular distributions near a surface , approximated the incident distribution through the use of an unknown constant ( which was  same as the unknown slip coefficient ) and then employed a conservation law with respect to stress to obtain the value of the unknown constants This problem is equivalent to the Milne problem of radiative transfer . The equation of radiative transfer in Milne problem consists of the integro-differential equation and boundary conditions relating  to the description of particle (photons, neutrons, molecules ) distribution in a semi infinite background medium where the number of particles is conserved in a collision with the background particles and where no  particles are incident on the free surface.  The form of the boundary condition at infinity is known from the asymptotic solution of the transport equation . It contains an undetermined constant and  that is related to the extrapolation distance  (where the particle concentration appears to vanish ) and is determined as a part of the solution of the problem .Here the ratio of the second moment to the first moment is very important to determine such unknown slip coefficient .

Apart from the slowing down of the neutrons in material medium, the neutrons also diffuse in the medium as a result of the scattering collisions Because of low density of neutrons in the medium the collisions between the neutrons are rare so we have to consider the collisions of the neutrons with the nuclei of material which may be regarded as stationary. A typical neutron trajectory consists of a number of short straight path elements . These are the scattering free paths. The average of these is called the mean free path . The neutron flux does not vanish at the boundary of the medium instead it vanishes at the linear extrapolation distance away from the medium. This linear extrapolation distance is of considerable importance on the choice of the thickness of the reflector for the nuclear reactor so that proper shielding is made to nuclear reactor to keep the scientists away from radiation hazards at the time of performing the experiments with high energy particles outside the shielding zone . This extrapolation distance in semi-infinite plane parallel conservative isotropic scattering atmosphere is the ratio of the second moment to the first moment of the H-function in radiative transfer .

### a) for non –conservative cases :

In equation (15a) , we shall use $U(u) = c/2$ , $0 \le c \le 1$ , to determine the form and numerics for the moments of H- function in non conservative, isotropic scattering cases as follows :

$$H(z) = A^{-1/2} ((1+z) / (1+kz)) \exp(-\int_0^1 \theta(t) \, dt / (t+z)) \, , \, (26a)$$

where

$$\theta(t) = \pi^{-1} \tan^{-1}(\pi t \, c/2 / T_0(t)) , \quad (26b)$$

$$A = D/k^2 ; \quad (27c)$$

$$D = 1 - c ; \quad (28)$$

$k$ , $(0 \le k < 1)$ , the positive zero of the even function $T(z)$

$$c = 2k / \log[(1+k)/(1-k)] \quad (29\,a)$$

$$T_0(x) = 1 - \tfrac{1}{2} x \, c \, \log((1+x)/(1-x)) , \quad (29b)$$

In non conservative cases ( $U_0 < 1/2$ ),for large $z$ , we may see $H(z)$ from equation (15a) as

$$H(z) = H_0 - H_{-1} / z + H_{-2} / z^2 - H_{-3} / z^3 + H_{-4} / z^4 \ldots \infty \, , \, (30)$$

where $H_0$ , $H_{-1}$, $H_{-2}$, $H_{-3}$, $H_{-4}$ ............. are unknown constants .

In non conservative cases , for large z , following equation (15a)  we can write

$$\int_0^1 \theta(t)\, dt / (t+z) = \theta_0 / z - \theta_1 / z^2 + \theta_2 / z^3 - \theta_3 / z^4 -- , (31)$$

$$\exp\left(- \int_0^1 \theta(t)\, dt / (t+z)\right) = \exp(- \theta_0 / z + \theta_1 / z^2 - \theta_2 / z^3 + \theta_3 / z^4 - -\infty )$$

$$= 1 - s_{-1} / z + s_{-2} / z^2 - s_{-3} / z^3 + s_{-4} / z^4 - - \infty , (32)$$

$$(1 + z) / (1 + kz) = k^{-1} - \epsilon_2 / z + \epsilon_3 / z^2 - \epsilon_4 / z^4 - -\infty , (33)$$

where

$$s_{-1} = \theta_0 , \tag{34a}$$
$$s_{-2} = \theta_0^2 / 2 + \theta_1 , \tag{34b}$$
$$s_{-3} = \theta_2 + \theta_0 \theta_1 + \theta_0^3 / 6 , \tag{34c}$$
$$s_{-4} = \theta_3 + \theta_0 \theta_2 + \theta_1^2 / 2 + \theta_0^2 \, \theta_1 / 2 , \quad \text{etc} \tag{34d}$$

$$\epsilon_r = ( k^{-r} - k^{-(r-1)} ) , r = 2,3,4, ------ , \tag{34e}$$

$$\theta_r = \int_0^1 \theta(x)\, x^r \, d x , \quad r = 0,1,2,3 , \tag{34f}$$

$\theta_r$'s  are called the moments of  $\theta(t)$  function.

Using equations (31,32,33) to the right side and using equation (30) to the left side of equation (26a) and equating like powers of $z^{-r}$, r=0,1,2,3 …from both sides of equation (26a) for large values of z , we shall get , the unknowns   $H_{-r}$, r=0,1,2,3,4……….as follows:

$$H_0 = A^{-1/2} k^{-1} ; \tag{3 5a}$$

$$H_{-1} = A^{-1/2} ( s_{-1} k^{-1} + \epsilon_2 ) ; \tag{35b}$$

$$H_{-2} = A^{-1/2} ( s_{-2} k^{-1} + s_{-1} \epsilon_2 + \epsilon_3 ) ; \tag{35c}$$

$$H_{-3} = A^{-1/2} ( s_{-3} k^{-1} + s_{-2} \epsilon_2 + s_{-1} \epsilon_3 + \epsilon_4 ) ; \tag{35d}$$

$$H_{-4} = A^{-1/2} ( s_{-4} k^{-1} + s_{-3} \epsilon_2 + s_{-2} \epsilon_3 + s_{-1} \epsilon_4 + \epsilon_5 ) ; \text{ etc} \tag{35e}$$

We can determine  $1/H(z)$ ,in non conservative cases  $U_0 < \frac{1}{2}$ , from equation (30) for large z as

$$1/H(z) = p_0 + p_{-1} / z + p_{-2} / z^2 + p_{-3} / z^3 + ---------- ; (36)$$

We can use for large z , equation (30) with equation (36)  to

$$H(z) \ ( \ H(z) \ )^{-1} = 1 \qquad ; \qquad\qquad\qquad\qquad (37)$$

to get

$$p_0 \ = 1/ \ H_0 \ ; \qquad\qquad\qquad (38a)$$
$$p_{-1} = p_0 \ H_{-1}/ \ H_0 \ ; \qquad\qquad (38b)$$
$$p_{-2} = (p_{-1} \ H_{-1} - p_0 \ H_{-2} \ )/H_0 \ ; \qquad (38c)$$

For large z ,  from equation (3) , we get for large z ,

$$1 / \ H(z) \ = 1- \ \alpha_0 + \alpha_1 \ / \ z - \alpha_2 \ / \ z^2 + \alpha_3 \ / \ z^3 - -- \ ; \qquad\qquad ( \ 39)$$

where $\quad \alpha_r = \frac{1}{2} c \ \int_0^1 \ x^r \ H(x) \, d \ x \ , \ r = 0,1,2,3 \ -- \ ; \ (40)$

$$\alpha_0 = \frac{1}{2} c \ \int_0^1 \ H(x) \, d \ x \ = \ 1 - ( \ 1 - c \ )^{1/2} \qquad ; \ (41)$$

$\alpha_r$'s   are called the weighted  moments of H- function.

From equation (36 ) and  (39)     we get  , the weighted  moments  of  H(z)  in  non conservative cases as ,

$$\alpha_0 = \ 1 - p_0 \ ; \ \alpha_1 = \ p_{-1} \ ; \ \alpha_2 = \ - \ p_{-2} \ ; \ \alpha_3 = \ p_{-3} \quad etc. \qquad ; \ (43)$$

Therefore the zeroth , first and second  moments of H-functions   are determined  as follows from  :

$$m_0 = 2\alpha_0/c \ ; \ \ m_1 = 2 \ \alpha_1/c \quad ; \ \ m_2 = 2 \ \alpha_2 /c \ ; \qquad\qquad (44)$$

where $\qquad m_r \ , \ r = 0,1,2 \quad$ are called the  moments of H- function.

In  non-conservative case , we therefore    outline   the calculation principle  for determination of  moments of H-functions as follows:

i)    At first , we shall    calculate   moments  of $\theta$-functions i.e.  $\theta_0$ , $\theta_1$ , $\theta_2$ , $\theta_3$ …….from the expression (26b) of $\theta$ (t) by way of integration using Simpson's one third rule ; ii)   after calculation of  $\theta_0$ , $\theta_1$ , $\theta_2$ , $\theta_3$  we shall determine $s_{-1}$ , $s_{-2}$ , $s_{-3}$ , $s_{-4}$ ,  from equation (34) ;iii)  after calculation of  $s_{-1}$ , $s_{-2}$ , $s_{-3}$ , $s_{-4}$ , ………. , we shall   determine $\epsilon_2$ , $\epsilon_3$ , $\epsilon_4$ …..from equation (34e) ,using the determined value of k from equation (29a) by iteration ; iv)  we shall   determine  $H_0$  , $H_{-1}$ , $H_{-2}$ , $H_{-3}$  ……. from equation (35a,35b,35c) : we shall determine $p_0$ , $p_{-1}$ , $p_{-2}$ , ……. from equation (38a,38b,38c) ; v)   after determination of $p_0$ , $p_{-1}$ , $p_{-2}$ , … we shall determine $\alpha_0$ , $\alpha_1$ , $\alpha_2$ , ……. from equation (43) ;vi)   thereafter $m_0, m_1, m_2$ , the zeroth, first and second  moments of H – function  respectively will be  determined  from equation (44) .

**b) for conservative cases :**

In  conservative case , we follow the same procedure of  non conservative  cases as follows:

for conservative case( $U_0$ =1/2 ) we shall use H(z) from equation (16a)  as

$$H(z) = C^{-1/2} (1+z) \exp\left(-\int_0^1 \theta(t) \, dt \, / \, (t+z)\right) \, , \qquad (45)$$

$$\theta(t) = (\pi)^{-1} \tan^{-1}\left(\tfrac{1}{2} \pi t \, / \, T_0(t)\right) ; \quad (46a)$$

$$C = 2 U_2 \; ; \qquad\qquad\qquad (46b)$$

$$U_2 = \int_0^1 x^2 U(x) \, dx \; , \qquad\qquad (46c)$$

$$T_0(x) = 1 - \tfrac{1}{2} \, x \, \log\left(( 1+x) /(1-x) \right)), \qquad (47)$$

For large  z , in conservative case , ($U_0 = \tfrac{1}{2}$) , we may get  H(z) after examination of equation (16a) as

$$H(z) = h_1 z + h_0 + h_{-1} / z + h_{-2} / z^2 + h_{-3} / z^3 + ---- \infty \; ; \; (48)$$

where $h_1$ , $h_0$ , $h_{-1}$ , $h_{-2}$ , ……… are unknown constants .

In  conservative cases , for large z , following Das[35], we can write

$$\int_0^1 \theta(t) \, dt \, / \, (t+z) = \theta_0 \, / z \, - \, \theta_1 \, / z^2 \, + \, \theta_2 / z^3 \, - \theta_3 / z^4 - - ; (49)$$

$$\exp\left(-\int_0^1 \theta(t) \, dt \, / \, (t+z)\right) = \exp\left(-\theta_0 \, / z \, + \, \theta_1 \, / z^2 \, - \theta_2 / z^3 \, + \theta_3 / z^4 - -\infty\right)$$

$$= 1 - s_{-1} / z + s_{-2} / z^2 \, - s_{-3} / z^3 + s_{-4} \, / z^4 \, - - \infty \quad ; (50)$$

where

$$s_{-1} = \theta_0 \; ; \qquad\qquad\qquad (51a)$$
$$s_{-2} = \theta_0^2 / 2 \, + \theta_1 \; ; \qquad\qquad (51b)$$
$$s_{-3} = \theta_2 + \theta_0 \theta_1 + \theta_0^3 / 6 \; ; \qquad (51c)$$
$$s_{-4} = \theta_3 + \theta_0 \theta_2 + \theta_1^2 / 2 + \theta_0^2 \, \theta_1 / 2 \; ; \quad \text{etc} \; ; (51d)$$

$$\theta_r = \int_0^1 \theta(x) \, x^r \, dx \; , \; r = \; 0,1,2,3 \ldots ; (51f)$$

$\theta_r$'s are called the moments of $\theta(t)$ function.

Using equations (49,50) to the right side and using equation (48) to the left side of equation (45) and equating like powers of $z^{-r}$, r=0,1,2,3…from both sides of equation (45) for large values of z, we shall get, the unknowns $h_{-r}$, r=0,1,2,3,4……….as follows

$h_1 = C^{-1/2}$ :                                             (52a)

$h_0 = (1 - s_{-1})\, h_1$  :                                   (52b)

$h_{-1} = (s_{-2} - s_{-1})\, h_1$    ;                         (52c)

$h_{-2} = (s_{-3} - s_{-2})\, h_1$    ;                         (52d)

$h_{-3} = (s_{-4} - s_{-3})\, h_1$    ;                         (52e)

$h_{-4} = (s_{-5} - s_{-4})\, h_1$    ;                         (52f)

where $s_{-1}$, $s_{-2}$, $s_{-3}$ .... are given by equations (51a,b,c,d) .

In conservative case, for large z, we shall get from equation (48)

$1/H(z) = n_{-1}/z + n_{-2}/z^2 + n_{-3}/z^3 + ----- \infty$ ,                  (53)
where

$n_{-1} = 1/h_1$;                                             (54a)

$n_{-2} = -\, h_0/h_1^2$ ;                                    (54b)

etc

For large z , from equation (3) , we get for large z , in conservative case ,

$1/H(z) = 1 - \alpha_0 + \alpha_1/z - \alpha_2/z^2 + \alpha_3/z^3 - --$ ;        ( 55)

where  $\alpha_r = \tfrac{1}{2} \displaystyle\int_0^1 x^r\, H(x)\, dx$ ,  r = 0,1,2,3 - - ;   (56a)

$\alpha_0 = \tfrac{1}{2} \displaystyle\int_0^1 H(x)\, dx = 1$         ; (56b)

$\alpha_r$ are called the weighted moments of H- functions.

From equation (55) and (53)  we get , the weighted moments of H(z) in conservative cases as ,

$\alpha_0 = 1$ ; $\alpha_1 = n_{-1}$ ; $\alpha_2 = - n_{-2}$ ; $\alpha_3 = n_{-3}$  etc.      ; (57)

Therefore the zeroth , first and second moments of H-functions are determined as follows from :

$m_0 = 2\alpha_0$ ;  $m_1 = 2\alpha_1$ ;  $m_2 = 2\alpha_2$ ;                      (58)

where   $m_r$ , r=0,1,2 are called the moments of H- functions.

 In conservative case , we therefore outline  the calculation principle of moments of H-functions as follows:

i)At first , we shall have  to calculate  moments of $\theta$-functions i.e  $\theta_0$ , $\theta_1$ , $\theta_2$ ,  $\theta_3$ …….from the expression (46a) of $\theta$ (t) by way of integration using Simpson's one   third rule ;ii) after calculation of  $\theta_0$ , $\theta_1$ , $\theta_2$ , $\theta_3$  we shall have to determine $s_{-1}$ , $s_{-2}$ , $s_{-3}$ , $s_{-4}$ , from equation (51a,b,c ) ;iii) we shall have  thereafter, to determine  $h_0$ , $h_1$ , $h_{-1}$ , $h_{-2}$ , $h_{-3}$ ……. from equation (52a,b,c)  ;iv)  we shall have thereafter , to determine  $n_{-1}$ , $n_{-2}$ , etc ……. from equation (54a,54b) ;  v) after determination of   $n_{-1}$ , $n_{-2}$ ,  … we shall be able to determine  $\alpha_0$ , $\alpha_1$ , $\alpha_2$ ,  ……. from equation (57) ;

vi)  thereafter $m_0$, $m_1$, $m_2$ , the zeroth, first and second   moments  of  H – function respectively will  be  determined  from equation (58) .

The extrapolation distance for non conservative and conservative cases i. e the ratio of the second moment to the first moment of H- function  are to be  determined  also and placed in a table .

### c) for **Extrapolated Distance:**

The comment of Loyalka and Naz  [7]  in determination of the extrapolation distance from the equation (9) instead of equation (59)

$$z_0 = 1 - \pi^{-1} \int_0^1 \tan^{-1}( \pi u/ (2\, T_0(u)) )\, d u \quad ; (59)$$

$$= 1 - \theta_0 \; ; \qquad\qquad (60)$$

is  that equation (9) is  a bit more efficient than equation (59) .This comment  may be ruled out   because  the ratio of the second moment to the first moment of H- function  in conservative case in our procedure  becomes the  same equation (59) and also equation (60) . Here  $\theta_0$ , the  zeroth moment  of $\theta$(x) is calculated in an easy and efficient way by use of Simpson's one third rule where  the description of $\theta$(x)  is  in equation (46a) read with  the equations ( 72 and 73 ) . The  extrapolation distance calculated form our methodology in conservative case ( this  is the maximum value of  all the ratios of second moment to first moment  in both non-conservative and conservative cases ) comes  for

$$\theta_0 = 0.28\ 955\ 391\ 268 \; ; \qquad\qquad (61)$$

to

$$z_0 = m_2/m_1 = 0.71\ 044\ 608\ 732 \; , \qquad\qquad (62)$$

which  matches with the result of  Plazeck and Siedel [10]  e.g equation  (10) and the result of Loyalka and Naz [7]up to eight places of decimals .

## B) Zero of Dispersion function ( pole of H-function)
## and numerics there for :

Following Ivanov [25], it can be said that in monochromatic scattering, the structure of radiation field depends in an essential way on  k , $0 \leq k \leq 1$, the positive root of the transcendental dispersion function T(z) . Far away from the source the radiation field is entirely determined by the value of k . The quantity  k is a monotonic function of c ,It is varying from k=1 at c=0 to k=0 at c=1. The reciprocal of this  k is called the diffusion length. It is important to emphasize that the diffusion length differs significantly   from mean free path of a photon only when absorption is relatively unimportant in comparison with scattering. That is why  a correct table for positive zero for T(z) is essential .

     Here  k is also used  to determine the pole of H(z)  in the complex plane cut along (-1,0)  and the accurate value  of this k  is required to use  in this paper  for numerical evaluation of H(u) , H$^{/}$(u) and H$^{//}$(u)  from their new forms.

In isotropic monochromatic scattering , we shall take U(u) = c/2 and k is determined as follows:

 Here k is  the positive root of the dispersion function T(z)  =0 ( e.g  equation (4) ) for the case U(u) =c/2. Hence k ,  $0 \leq k \leq 1$ is to be determined  from

$$c = 2 k / \log [(1+k) / (1-k) ] \ , \ \ 0 \leq c \leq 1 \ . \quad (63 )$$

We have ascertained the sufficient condition  of convergence of the root of the equation (63 ) .The asymptotic expansion of equation  ( 63) for small c gives

$$k= 1- 2\exp(-2/c)( \ 1 + (4-c) \exp(-2/c) \ +\text{------------} \ \ ) ; (64)$$

and for c close to unity  gives

$$k= (3(1-c) \ )^{\frac{1}{2}} ( \ 1- 2(1-c)/5 - \ \ \text{---------------------} \ ) : (65)$$

It is thus evident that in the extreme case of very weak absorption ( c$\rightarrow$1)

   the diffusion length is approximately =  $(3(1-c) \ )^{-\frac{1}{2}} =1/k. \ \ ;(66)$

In cases  $0 <c<1$, the equation ( 63 )  is converted to the following form  to make it useful for application of the method of iteration  :

$$k=1-2/(1+\exp(2k/c)). \quad\quad\quad (67)$$

We shall first take the initial guess of k  from equation  (63)

$$k = (3c(1-c) \ )^{\frac{1}{2}} , \ 0 <k<1 \quad\quad\quad (68)$$

   We shall, there after use the method of iteration to the equivalent function (67) and  find the value of k  correct  to ninth  places of decimals and those are compared with the roots obtained  from  Chandrasekhar ( [1] , page 19) and Ivanov ([25],page 105)as follows in a table no- 1:

### C )Differential properties of H-function

### and numerics there for :

The theory of light scattering in optically large spheres developed by Van de Hulst [5,6] has called a better understanding  and an increased knowledge of the first and second derivatives of H-functions , since the first two derivatives of the H-function with respect to direction parameter play an important role in the theory.

H- function and its derivatives are also used to model the reflection of light by particulate surface material of celestial bodies such as planets , moons, and asteroids. Those are also used in studies of radiation transport in spherical clouds.Van de Hulst [5] found an expression in closed form for the spherical reflection function of a homogeneous sphere with isotropic scattering if the radiation field is spherically symmetric. .On developing the asymptotic theory (the dominant deviation for optically large spheres from the well known theory for spheres with infinite large optical diameter ), he introduced a linear differential operator  of the second order on H-function . For instance ,this gives  an the asymptotic expression for the radiance   leaving the cloud under the angle  with the normal if the cloud is exposed to the uniform incidence radiance one. One advantage of having tables of $H^{/}(u)$, first derivative and $H^{//}(u)$ , second derivative along with H(u) is that it enables rapid and simple interpolation which is frequently occurring problem in radiative transfer .The differential properties of H- functions may be used to understand the angular dependence of the radiation emerging from a star or a planet .

Little attention has been paid in the literature to the derivatives of H-functions and related functions   with respect to the directional parameters.For isotropic scattering the first such derivative generally has a singularity  at u=0.The first of such , the  paper of Hovenier ,Mee and Heer[31] is  to establish certain differential properties of H-functions and related functions , in particular for isotropic scattering  and to provide some computational procedures along with tables of numerical results. They have considered the Chandrasekhar's H function  as in  equation (1) for U(u)= c/2 for determination of H-function  and differentiated the non linear integral equation(1) twice   to get the expression of $H^{/}(u)$ and $H^{//}(u)$   .They evaluated those numerically with the Gaussian interpolation   and  evaluated those $H^{/}(u)$ and $H^{//}(u)$  for further  numerical interpolation if required .  The expression of $H^{/}(u)$ and $H^{//}(u)$   used by them are as follows :

$$H^{/}(u) = \tfrac{1}{2} c \ H(u)^2 \int_0^1 \ x \ H(x) \ d \ x \ / \ (x + u)^2 \ ; \qquad\qquad (69)$$

$$H^{//}(u) = 2(\,H^{/}(u)\,)^2 \,/\, H(u) - c\,H(u)^2 \int_0^1 x\,H(x)\,d\,x \,/\,(x + u)^3 \ \ ;\,(70\ )$$

In this paper we shall consider the new expression of H(u) (cf. equation (15a,16a) herein)as obtained from Das [35,36] for numerical evaluation both for non conservative and conservative cases respectively and new expressions for $H^{/}(u)$ and $H^{//}(u)$ determined from those are outlined below for numerical evaluation :

### D) New expression of $H^{/}(u)$ and $H^{//}(u)$ in non conservative cases :

Equation (15a) , on differentiation , with respect to u , gives $H^{/}(u)$,first derivative of H(u) for non conservative cases as

$$H^{/}(u) \ = H(u)\,[\ 1/(1+u) - k/(1+k\,u) + \int_0^1 \theta\,(x)\ d\,x\ /\ (\,x + u)^2\,]\quad;\quad(71\ )$$

Following the suggestion of Williams [8], we can write modify $\theta\,(x)$ as

$$\theta\,(x) = 1/\,2 - \theta_1\,(x)\ \ ;\qquad\qquad\qquad(72)$$

where

$$\theta_1\,(x) = \pi^{-1}\ \tan^{-1}(\ T_0\,(x)\,/\,\pi\,x\ U(x)\ )\ \ ;\qquad(73)$$

Equation (71) with equations (72) and (73) takes the form of $H^{/}(u)$,first derivative of H(u) which is suitable for numerical evaluation in non conservative cases as

$$H^{/}(u) \ = H(u)\,[.5\,(1/(1+u) + 1/u\,) - k/(1+ku) - \int_0^1 \theta_1\,(x\,)\ d\,x\ /\ (\,x + u)^2\,]\quad;\quad(74)$$

We can differentiate equation (15a) twice with respect to u to get form of $H^{//}(u)$ , second derivative of H(u) which is also suitable for numerical evaluation for non conservative cases as

$$H^{//}(u) \ = [\ H^{/}(u)\,]^2\,/\,H(u)$$

$$-\,H(u)\,[\ 1/(1+u)^2 - k^2/(1+k\,u)^2 + 2\int_0^1 \theta\,(x)\ d\,x\ /\ (\,x + u)^3\,]\quad.\quad(75)$$

### E)New expression of $H^{/}(u)$ and $H^{//}(u)$ in conservative cases :

Equation (16a) , on differentiation , with respect to u , gives $H^{/}(u)$,first derivative of H(u) for conservative cases as

$$H^{/}(u) \ = H(u)\,[\ 1/(1+u) + \int_0^1 \theta\,(x)\ d\,x\ /\ (\,x + u)^2\,]\quad;\qquad(76\ )$$

Equation (76) with equations (72) and (73) takes the form of $H^/(u)$, first derivative of H(u) which is suitable for numerical evaluation for conservative cases as

$$H^/(u) = H(u) [.5 (1/(1+u) + 1/u) - \int_0^1 \theta_1(x) \, dx / (x+u)^2] \quad ; \quad (77)$$

We can differentiate equation (16a) twice with respect to u to get form of $H^{//}(u)$, second derivative of H(u) which is also suitable for numerical evaluation for conservative cases as

$$H^{//}(u) = [H^/(u)]^2 / H(u)$$
$$- H(u) [1/(1+u)^2 + 2 \int_0^1 \theta(x) \, dx / (x+u)^3] \quad ; \quad (78)$$

where θ(x) will be given by equations (72) and (73) for c=1.

## F) Steps for determination of H(u) , H$^/$(u) and H$^{//}$(u) :

For numerical evaluation of H(u) , H $^/$(u) and H $^{//}$(u) we shall take the following steps :

### ( i) for non conservative cases U(u) < ½ :

i)We shall at first determine the k from equation (67) , the zeros of dispersion function for U(u) = c/2 , by iteration for a particular c , albedo for single scattering ;ii)we shall use the modified expression for θ(x) as opined by Williams [8] and stated in equations ( 72& 73 ) ;iii)We shall determine the integrals in H(u) (cf. equation 26a); for a fixed u and c by Simpson's one third rule ;iv)We shall determine the value of A from equations ( 27c ,28) ; v)We shall determine the integrals in H$^/$(u) )(cf. equation ( 74 ); for a fixed u and c by Simpson's one third rule ;vi)We shall determine the integrals in H$^{//}$(u) )(cf.equation ( 75)); for a fixed u and c by Simpson's one third rule ; then use of aforesaid results in equations (26a) , (74 ) and ( 75 ) will give us the numerical result for H(u) , H$^/$(u) and H$^{//}$(u) for a fixed c with different values of the direction parameter u .

### (ii) for conservative cases U(u) = ½ :

: i)We shall use the expression for θ(x) as opined by Williams [8] and stated in equations ( 72 and 73 ) ;ii)We shall determine the integrals in H(u)(eg equation (45)); for a fixed u by Simpson's one third rule ; iii)We shall determine the value of C from equations (46b and 46c) ; iv)We shall determine the integrals in H$^/$(u) )(cf. equation ( 77 ); for a fixed u by Simpson's one third rule ;v)We shall determine the integrals in H$^{//}$(u) )(cf. equation (78 )); for a fixed u by Simpson's one third rule ;then use of aforesaid results in equations (45) , ( 77 ) and ( 78 ) will give us the numerical result for H(u) , H$^/$(u) and $H''(u)$  for  a fixed c  with different values of the direction parameter u  in conservative cases .

## G) Linear interpolation of H(u) using $H'(u)$ and $H''(u)$ and Numerics there for :

For fixed c , $0 < c \leq 1$  ,  H(u)  is monotonically  increasing in u where $0 \leq u \leq 1$ from H(0) = 1 to  H(1) and for fixed  u , $0 \leq u \leq 1$ ,  H(u) is monotonically increasing   with c where $0 < c \leq 1$ from  H(u)=1  to a value greater than  one.

For fixed c , $0 < c \leq 1$  ,  $H'(u)$  is monotonically  decreasing in u where $0 \leq u \leq 1$ from infinity  to  $H'(1)$, a positive quantity and for fixed  u , $0 \leq u \leq 1$ ,  $H'(u)$ is monotonically increasing   with c where $0 < c \leq 1$ from a value grater than zero but less than one  to a value greater than one .

For   fixed c , $0 < c \leq 1$  ,  $H''(u)$  is monotonically  increasing  with  u, $0 \leq u \leq 1$ from minus infinity to  $H''(1)$, a negative quantity i.e  $H''(u)$ is less than zero in the entire range of u and for fixed  u , $0 \leq u \leq 1$ ,  $H''(u)$ is monotonically decreasing  with c where  $0 < c \leq 1$ from a negative value   to a further negative value i.e  $H''(u)$ is less than zero in the entire range of c.

If c is close to zero , H(u) ,$H'(u)$ and $H''(u)$ are analytic function of  c near c=0  and those functions  may be expressed  in a power series  in c as

$$H (u)  = 1+ c/2\ u\ \log (1+1/u) + 0(c^2) ; \qquad (79)$$
$$H'(u) =\ c/2\ \{ \log(1+1/u) -1/(1+u)\ )\ +0(c^2) ; \qquad (80)$$
$$H''(u)  = - [\ c/ (2\ u\ (1+u)^2 )\ ] +0(c^2 ) ; \qquad (81)$$

If c is near 1 ,  we can have the expansion

$$H(u) = H_c(u) -\ 3^{1/2}\ t\ u\ H_c(u) + 0(t^2)\ ; \qquad (82)$$

$$H'(u) = H'_c(u) - 3^{1/2}\ t\ (\ H_c(u) + u\ H'_c(u)\ ) +0(t^2)\ ; \qquad (83)$$

$$H''(u) = H''_c(u) - 3^{1/2}\ t\ (2\ H'_c(u) + u\ H''_c(u)\ ) +0(t^2)\ ; \qquad (84)$$

where $H_c(u)$ $H_c'(u)$ ,  $H_c''(u)$ are   the H- function , its first and second derivative  for conservative scattering  (c=1) and t = $(1-c)^{1/2}$ and those are the equations (5) ,(77)   and (78) respectively.

One advantage of having tables of $H'(u)$ and $H''(u)$ along with H(u) is that it enables rapid and simple interpolation of H(u) , which is of  a frequently occurring problems in radiative transfer . We  can write for u >0 and 0< h < 1

$$H (u +h) = H(u) + h\ H'(u) + \tfrac{1}{2!}\ h^2\ H''(u) +----------------- .(85)$$

Use  of  equation (85)   will give the interpolated  H(u) and that value may be compared with the tabulated value  in tables nos 2 to 8.

**Table for zero of T(z)( pole of H(z) )**

**( determined**

**from equation (67)**

**and its comparison with available results )**

**(table -1)**

| c | k determined by us from equation(67 ) | k obtained from from equation( 67 ) (cf. Chandrasekhar[1], page19) | k obtained by from equation(67 ) (cf. Ivanov [25], page105) |
|---|---|---|---|
| 0 | 1 | 1 | 1 |
| 0.1 | 0.999 999 995 88 | | |
| 0.2 | 0.999 909 121 72 | 0.99991 | |
| 0.3 | 0.997 413 816 89 | 0.99741 | 0.9974 |
| 0.4 | 0.985 623 871 63 | 0.98562 | 0.9856 |
| 0.5 | 0.957 504 024 08 | 0.9575 | 0.9575 |
| 0.6 | 0.907 332 316 64 | 0.90733 | 0.9073 |
| 0.7 | 0.828 634 798 63 | 0.82864 | 0.8286 |
| 0.8 | 0.710 411 783 48 | 0.71041 | 0.7104 |
| 0.9 | 0.525 429 512 62 | 0.52543 | 0.5254 |
| 0.925 | 0.459 925 287 65 | 0.45993 | |
| 0.95 | 0.379 485 206 59 | 0.37948 | |
| 0.975 | 0.271 110 860 70 | 0.27111 | |
| 1 | 0 | 0 | 0 |

# Tables of  H(u) , H′(u)  and H″(u)
## (determined
## from
## equations (15a,74 and 75))
## (table −2 )

| u | c= 0.1 H(u) | H′(u) | H″(u) | c= 0.2 H(u) | H′(u) | H″(u) |
|---|---|---|---|---|---|---|
| 0.0 | 1.00 000 000 000 | + ∞ | - ∞ | 1.00 000 000 000 | + ∞ | - ∞ |
| 0.05 | 1.00 780 896 944 | 0.10 855 453 668 | - 0.91 118 360 895 | 1.01 605 299 583 | 0.22 586 684 758 | -1.82 893 782 344 |
| 0.1 | 1.01 237 809 580 | 0.07 811 661 199 | -0.41 923 593 934 | 1.02 561 781 551 | 0.16 449 877 653 | -0.85 005 119 927 |
| 0.15 | 1.01 584 150 461 | 0.06 174 135 724 | -0.25 794 903 290 | 1.03 294 308 025 | 0.13 115 230 786 | -0.52 784 709 160 |
| 0.2 | 1.01 864 407 200 | 0.05 101 351 996 | -0.17 901 782 488 | 1.03 891 699 513 | 0.10 911 426 604 | -0.36 929 967 057 |
| 0.25 | 1.02 099 253 430 | 0.04 330 762929 | -0.13 285 316 169 | 1.04 395 458 583 | 0.09 316 261 933 | -0.27 602 071 809 |
| 0.3 | 1.02 300 555 994 | 0.03 746 169 298 | -0.10 294 826 566 | 1.04 829 552 044 | 0.08 097 946 845 | -0.21 523 897 655 |
| 0.35 | 1.02 475 933 136 | 0.03 286 124 278 | -0.08 224 313 516 | 1 .05 209 457 762 | 0.07 133 458 388 | -0.17 291 618 972 |
| 0.4 | 1.02 630 632 664 | 0.02 914 351 417 | -0.06 721 848 269 | 1.05 545 898 990 | 0.06 349 869 849 | -0.14 203 894 436 |
| 0.45 | 1.02 768 450 560 | 0.02 607 757 982 | -0.05 592 743 332 | 1.05 846 674 616 | 0.05 700 559 052 | -0.11 871 694 622 |
| 0.5 | 1.02 892 233 441 | 0.02 350 813 205 | -0.04 720 835 027 | 1.06 117 662 957 | 0.05 154 034 237 | -0.10 062 193 515 |
| 0.55 | 1.03 004 176 197 | 0.02 132 624 809 | -0.04 032 708 225 | 1.06 363 418 173 | 0.04 688 115 019 | -0.08 687 769 877 |
| 0.6 | 1.03 106 009 162 | 0.01 945 296 681 | -0.03 479 818 680 | 1.06 587 546 536 | 0.04 286 654 744 | -0.07 470 484 361 |
| 0.65 | 1.03 199 121 443 | 0.01 782 949 397 | -0.03 028 874 240 | 1.06 792 955 191 | 0.03 937 581 515 | -0.06 522 939 908 |
| 0.7 | 1.03 284 645 267 | 0.01 641 106 089 | -0.02 656 345 360 | 1.06 982 022 892 | 0.03 631 668 761 | -0.05 737 340 629 |
| 0.75 | 1.03 363 515 491 | 0.01 516 294 594 | --0.02 345 177 152 | 1.07 156 720 837 | 0.03 361 732 645 | -0.05 078 924 195 |
| 0.8 | 1.03 436 512 651 | 0.01 405 777 897 | -0.02 082 739 704 | 1.07318 700 422 | 0.03 122 090 575 | -0.04 521 862 121 |
| 0.85 | 1.03 504 294 820 | 0.01 307 368 526 | -0.01 859 503 451 | 1.07 469 358 357 | 0.02 908 185 523 | -0.04 046 605 180 |
| 0.9 | 1.03 567 421 604 | 0.01 219 296 883 | -0.01 668 163 327 | 1.07 609 885 903 | 0.02 716 319 083 | -0.03 638 117 629 |
| 0.95 | 1.03 626 372 531 | 0.01 140 115 981 | -0.01 503 038 036 | 1.07 741 306 716 | 0.02 543 457 800 | -0.03 284 670 834 |
| 1 | 1.03 681 561 345 | 0.01 068 631 272 | -0.01 359 653 392 | 1.07 864 506 335 | 0.02 387 090 108 | -0.02 977 000 787 |

### Tables of  H(u) , H$^{/}$(u)  and H$^{//}$(u)
### (determined
### from
### equations(15a,74 and 75))
### (table-3)

| u | H(u) | H$^{/}$(u) | H$^{//}$(u) | H(u) | H$^{/}$(u) | H$^{//}$(u) |
|---|---|---|---|---|---|---|
| | | c= 0.3 | | | c= 0.4 | |
| 0.0 | 1.00 000 000 000 | + ∞ | - ∞ | 1.00 000 000 000 | + ∞ | -∞ |
| 0.05 | 1.02 480 507 781 | 0.35 352 742 239 | -2.74 976 108 619 | 1.03 416 322 685 | 0.49 368 012 768 | -3.66 842 935 352 |
| 0.1 | 1.03 987 488 559 | 0.26 084 608 289 | -1.29 126 761 629 | 1.05 535 954 887 | 0.36 947 931 618 | -1.74 065 777 344 |
| 0.15 | 1.05 154 635 878 | 0.20 996 016 809 | -0.80 964 412 272 | 1.07 197 966 305 | 0.30 055 445 244 | -1.10 267 140 909 |
| 0.2 | 1.06 114 652 881 | 0.17 601 520 939 | -0.57 141 002 859 | 1.08 578 088 067 | 0.25 411 457 623 | -0.78 559 229 201 |
| 0.25 | 1.06 929 869 234 | 0.15 124 028 088 | -0.43 041 149 675 | 1.09 759 240 857 | 0.21 991 166 248 | -0.59 682 380 355 |
| 0.3 | 1.07 636 495 745 | 0.13 217 798 222 | -0.33 796 976 381 | 1.10 789 881 878 | 0.19 337 918 821 | -0.47 227 796 173 |
| 0.35 | 1.08 258 062 810 | 0.11 698 712 115 | -0.27 321 170 538 | 1 .11 701 701 130 | 0.17 207 843 794 | -0.38 446 900 280 |
| 0.4 | 1.08 810 972 310 | 0.10 457 191 592 | -0.22 569 065 366 | 1. 12 516 930 560 | 0.15 455 267 216 | -0.31 962 540 682 |
| 0.45 | 1.09 307 223 335 | 0.09 422 879 945 | -0.18 959 806 106 | 1.13 251 938 018 | 0.13 986 260 395 | -0.27 007 560 186 |
| 0.5 | 1.09 755 910 390 | 0.08 548 491 205 | -0.16 144 775 443 | 1.13 919 205 779 | 0.12 736 803 214 | -0.23 120 387 370 |
| 0.55 | 1.10 164 117 243 | 0.07 798 900 627 | -0.13 902 235 037 | 1.14 528 512 437 | 0.11 661 384 730 | -0.20 006 576 743 |
| 0.6 | 1.10 537 480 453 | 0.07 150 761 016 | -0.12 084 575 968 | 1.15 087 683 116 | 0.10 726 566 360 | -0.17 469 472 709 |
| 0.65 | 1.10 880 565 242 | 0.06 585 076 549 | -0.10 589 863 451 | 1.15 603 088 756 | 0.09 907 118 502 | -0.15 372 790 241 |
| 0.7 | 1.11 197 123 407 | 0.06 087 612 069 | -0.09 345 548 600 | 1.16 079 991 214 | 0.09 183 584 009 | -0.13 619 173 881 |
| 0.75 | 1.11 490 276 705 | 0.05 647 238 334 | -0.08 298 677 712 | 1.16 522 789 309 | 0.08 540 681 964 | -0.12 137 293 440 |
| 0.8 | 1.11 762 650 499 | 0.05 255 117 167 | -0.07 409 763 093 | 1.16 935 198 813 | 0.07 966 222 591 | -0.10 873 753 982 |
| 0.85 | 1.12 016 473 296 | 0.04 904 133 316 | -0.06 648 811 857 | 1.17 320 387 024 | 0.07 450 351 181 | -0.09 787 841 684 |
| 0.9 | 1.12 253 652 281 | 0.04 588 491 824 | -0.05 992 676 099 | 1.17 681 075 266 | 0.06 985 007 417 | -0.08 848 015 234 |
| 0.95 | 1.12 475 831 553 | 0.04 303 422 875 | -0.05 423 236 586 | 1.18 019 618 252 | 0.06 563 530 737 | -0.08 029 493 819 |
| 1 | 1.12 684 437 695 | 0.04 044 964 136 | -0.04 926 135 731 | 1.18 338 066 380 | 0.06 180 366 962 | -0.07 312 566 678 |

**Tables of H(u) , H$^{/}$(u)  and H$^{//}$(u)
(determined
from
equations(15a,74 & 75))
(table-4)**

| | | | c= 0.5 | | | | c= 0.6 | |
|---|---|---|---|---|---|---|---|---|
| u | H(u) | H$^{/}$(u) | H$^{//}$(u) | H(u) | H$^{/}$(u) | H$^{//}$(u) |
| 0.0 | 1.00 000 000 000 | +∞ | -∞ | 1.00 000 000 000 | +∞ | -∞ |
| 0.05 | 1.04 426 515 806 | 0.64 934 753 603 | -4.57 689 620 274 | 1.05 531 672 613 | 0.82 506 115 268 | -5.46 213 733 606 |
| 0.1 | 1.07 236 875 955 | 0.49 371 285 649 | -2.19 414 191 263 | 1.09 134 858 796 | 0.63 857 896 945 | -2.54 414 444 029 |
| 0.15 | 1.09 479 972 966 | 0.40 639 381 377 | -1.40 505 326 980 | 1.12 044 221 493 | 0.53 280 278 266 | -1.71 243 895 587 |
| 0.2 | 1.11 346 142 387 | 0.34 692 782 856 | -1.01 134 900 293 | 1.14 516 450 698 | 0.45 993 996 852 | ˙1.24 647 748 424 |
| 0.25 | 1.12 965 308 865 | 0.30 269 305 990 | -0.77 566 001 950 | 1.16 673 340 619 | 0.40 514 005 256 | -0.96 621 273 046 |
| 0.3 | 1.14 388 950 385 | 0.26 806 333 709 | -0.61 916 245 984 | 1.18 586 786 173 | 0.36 179 335 877 | -0.77 897 698 553 |
| 0.35 | 1 .15 656 870 841 | 0.24 002 828 769 | -0.50 808 155 006 | 1.20 304 393 028 | 0.32 636 190 124 | -0.64 516 123 171 |
| 0.4 | 1.16 797 184 530 | 0.21 678 416 616 | -0.42 549 280 851 | 1.21 860 018 021 | 0.29 672 175 309 | -0.54 494 523 941 |
| 0.45 | 1.17 830 731 727 | 0.19 716 339 709 | -0.36 195 865 031 | 1.23 278 941 185 | 0.27 149 381 757 | -0.46 728 080 691 |
| 0.5 | 1.18 773 512 842 | 0.18 036 662 263 | -0.31 179 058 293 | 1.24 580 715 523 | 0.24 973 013 986 | -0.40 550 469 570 |
| 0.55 | 1.19 638 144 842 | 0.16 582 284 272 | -0.27 135 091 669 | 1.25 780 872 990 | 0.23 075 041 776 | -0.35 534 968 676 |
| 0.6 | 1.20 434 787 955 | 0.15 311 040 744 | -0.23 820 287 429 | 1.26 892 011 488 | 0.21 404 970 323 | -0.31 395 100 016 |
| 0.65 | 1.21 171 764 229 | 0.14 190 953 144 | -0.21 065 203 342 | 1.27 924 522 001 | 0.19 924 297 769 | -0.27 931 094 455 |
| 0.7 | 1.21 855 986 525 | 0.13 197 234 010 | -0.18 748 351 837 | 1.28 887 094 298 | 0.18 603 012 140 | -0.24 999 240 622 |
| 0.75 | 1.22 493 265 486 | 0.12 310 315 499 | -0.16 780 375 096 | 1.29 787 079 999 | 0.17 417 387 089 | -0.22 493 436 579 |
| 0.8 | 1.23 088 534 982 | 0.11 514 509 956 | -0.15 094 110 212 | 1.30 630 760 221 | 0.16 347 912 367 | -0.20 333 609 393 |
| 0.85 | 1.23 646 021 345 | 0.10 797 072 593 | -0.13 638 136 791 | 1.31 423 547 309 | 0.15 379 193 442 | -0.18 458 190 127 |
| 0.9 | 1.24 169 372 811 | 0.10 147 529 510 | -0.12 372 450 266 | 1.32 170 139 798 | 0.14 498 160 652 | -0.16 819 064 504 |
| 0.95 | 1.24 661 760 151 | 0.09 557 185 669 | -0.11 265 489 707 | 1.32 874 643 421 | 0.13 693 988 497 | -0.15 378 099 691 |
| 1 | 1.25 125 956 002 | 0.09 018 757 951 | -0.10 292 047 074 | 1.33 540 666 947 | 0.12 957 561 195 | -0.14 104 697 297 |

**Tables of  H(u) , H$'$(u)  and H$''$(u)**
**(determined**
**from**
**equations(15a,74 & 75))**
**(table- 5)**

|  | c=0.7 | | | c=0.8 | | |
|---|---|---|---|---|---|---|
| u | H(u) | H$'$(u) | H$''$(u) | H(u) | H$'$(u) | H$''$(u) |
| 0.0 | 1.00 000 000 000 | +∞ | -∞ | 1.00 000 000 000 | +∞ | -∞ |
| 0.05 | 1.06 765 459 763 | 1.02 824 967 897 | -6.30 135 963 791 | 1.08 191 451 387 | 1.27 296 409 722 | -7.04 906 473 873 |
| 0.1 | 1.11 303 183 363 | 0.81 242 9/5 773 | -3.07 578 964 511 | 1.13 880 766 122 | 1.03 124 492 500 | -3.45 623 074 369 |
| 0.15 | 1.15 034 382 426 | 0.68 874 451 241 | -2.01 492 457 376 | 1.18 664 007 755 | 0.89 162 257 320 | -2.28 802 662 369 |
| 0.2 | 1.18 251 578 032 | 0.60 251 668 656 | -1.48 459 483 225 | 1.22 863 876 052 | 0.79 312 188 043 | -1.70 750 363 922 |
| 0.25 | 1.21 093 411 034 | 0.53 686 922 642 | -1.16 470 396 959 | 1.26 632 248 271 | 0.71 712 617 198 | -1.35 787 550 572 |
| 0.3 | 1.23 641 931 989 | 0.48 432 333 059 | -0.94 990 986 670 | 1.30 058 827 353 | 0.65 545 925 133 | -1.12 266 114 861 |
| 0.35 | 1 .25 951 740 611 | 0.44 088 479 902 | -0.79 539 489 691 | 1 .33 203 405 119 | 0.60 378 511 000 | -0.95 268  611 431 |
| 0.4 | 1.28 061 919 723 | 0.40 415 667 958 | -0.67 881 570 462 | 1.36 108 969 694 | 0.55 951 453 281 | -0.82 361 329 609 |
| 0.45 | 1.30 001 868 203 | 0.37 258 057 685 | -0.58 775 102 449 | 1.38 808 064 115 | 0.52 096 891 227 | -0.72 200 200 214 |
| 0.5 | 1.31 794 505 873 | 0.34 508 242 096 | -0.51 472 147 655 | 1.41 326 256 478 | 0.48 699 231 899 | -0.63 980 466 634 |
| 0.55 | 1.33 458 188 756 | 0.32 088 847 890 | -0.45 493 966 735 | 1.43 684 201 745 | 0.45 675 125 229 | -0.57 189 357 312 |
| 0.6 | 1.35 007 928 700 | 0.29 942 177 823 | -0.40 518 999 179 | 1.45 898 948 554 | 0.42 962 263 088 | -0.51 483 596 663 |
| 0.65 | 1.36 456 209 562 | 0.28 024 004 272 | -0.36 322 690 345 | 1.47 984 810 775 | 0.40 512 699 639 | -0.46 623 964 741 |
| 0.7 | 1.37 813 555 637 | 0.26 299 652 441 | -0.32 743 134 165 | 1.49 953 973 165 | 0.38 288 660 979 | -0.42 438 129 044 |
| 0.75 | 1.39 088 940 664 | 0.24 741 425 778 | -0.29 660 437 962 | 1.51 816 926 770 | 0.36 259 801 211 | -0.38 798 437 139 |
| 0.8 | 1.40 290 090 214 | 0.23 326 852 356 | -0.26 983 801 771 | 1.53 582 790 897 | 0.34 401 340 391 | -0.35 608 079 480 |
| 0.85 | 1.41 423 710 425 | 0.22 037 454 631 | -0.24 643 122 660 | 1.55 259 556 988 | 0.32 692 761 758 | -0.32 792 138 973 |
| 0.9 | 1.42 495 664 420 | 0.20 857 864 121 | -0.22 583 373 720 | 1.56 854 277 141 | 0.31 116 875 964 | -0.30 291 622 940 |
| 0.95 | 1.43 511 110 671 | 0.19 775 170 117 | -0.20 760 735 750 | 1.58 373 212 442 | 0.29 659 133 374 | -0.28 059 370 541 |
| 1 | 1.44 474 613 115 | 0.18 778 431 276 | -0.19 139 871 819 | 1.59 821 951 461 | 0.28 307 108 419 | -0.26 057 177 876 |

**Tables of  H(u) , H$'$(u)  and H$''$(u)**
**(determined**
**from**
**equations(15a,74 & 75))**
**(table-6)**

|  | c=0.9 | | | c=0.925 | | |
|---|---|---|---|---|---|---|
| u | H(u) | H$'$(u) | H$''$(u) | H(u) | H$'$(u) | H$''$(u) |
| 0.0 | 1.00 000 000 000 | +∞ | -∞ | 1.00 000 000 000 | +∞ | -∞ |
| 0.05 | 1.09 967 828 295 | 1.59 418 923 918 | -7.58 718 876 030 | 1.10 517 492 801 | 1.69 768 847 522 | -7.65 029 811 619 |
| 0.1 | 1.17 214 304 834 | 1.33 524 591 863 | -3.69 198 305 537 | 1.18 277 357 379 | 1.43 755 800 185 | -3.69 586 228 881 |
| 0.15 | 1.23 491 832 726 | 1.18 589 312 628 | -2.45 500 468 658 | 1.25 066 435 601 | 1.28 825 565 740 | -2.45 190 952 928 |
| 0.2 | 1.29 143 372 282 | 1.07 970 577 551 | -1.85 174 501 843 | 1.31 230 132 304 | 1.18 218 854 517 | -1.85 066 639 607 |
| 0.25 | 1.34 327 082 023 | 0.99 674 923 338 | -1.49 335 306 213 | 1.36 926 230 492 | 1.09 918 791 879 | -1.49 631 986 025 |
| 0.3 | 1.39 135 031 541 | 0.92 841 590 422 | -1.25 430 626 025 | 1.42 245 823 144 | 1.03 059 882 221 | -1.26 154 308 969 |
| 0.35 | 1 .43 628 033 816 | 0.87 021 547 531 | -1.08 224 643 583 | 1 .47 248 666 479 | 0.97 193 609 144 | -1.09 342 775 225 |
| 0.4 | 1.47 849 623 100 | 0.81 950 719 341 | -0.95 158 881 913 | 1.51 977 311 711 | 0.92 058 012 046 | -0.96 623 027 505 |
| 0.45 | 1.51 832 748 723 | 0.77 460 194 491 | -0.84 838 911 249 | 1.56 463 843 480 | 0.87 486 685 189 | -0.86 598 603 535 |
| 0.5 | 1.55 603 379 696 | 0.73 434 958 190 | -0.76 440 881 931 | 1.60 733 491 262 | 0.83 367 072 861 | -0.78 448 699 937 |
| 0.55 | 1.59 182 617 759 | 0.69 792 803 516 | -0.69 446 671 964 | 1.64 806 735 482 | 0.79 619 223 690 | -0.71 659 948 036 |
| 0.6 | 1.62 588 019 124 | 0.66 472 660 949 | -0.63 513 704 981 | 1.68 700 617 640 | 0.76 184 074 160 | -0.65 894 861 452 |
| 0.65 | 1.65 834 464 992 | 0.63 427 715 879 | -0.58 405 963 700 | 1.72 429 598 218 | 0.73 016 565 328 | -0.60 922 290 081 |
| 0.7 | 1.68 934 759 547 | 0.60 621 140 138 | -0.53 955 146 591 | 1.76 006 142 554 | 0.70 081 391 736 | -0.56 578 364 576 |
| 0.75 | 1.71 900 055 437 | 0.58 023 328 860 | -0.50 037 672 338 | 1.79 441 135 245 | 0.67 350 261 905 | -0.52 743 442 080 |
| 0.8 | 1.74 740 165 639 | 0.55 610 046 490 | -0.46 560 489 517 | 1.82 744 182 172 | 0.64 800 068 965 | -0.49 327 925 116 |
| 0.85 | 1.77 463 797 970 | 0.53 361 143 730 | -0.43 451 988 224 | 1 .85 923 836 359 | 0.62 411 630 982 | -0.46 263 210 320 |
| 0.9 | 1.80 078 735 342 | 0.51 259 645 384 | -0.40 656 001 414 | 1.88 987 770 709 | 0.60 168 800 010 | -0.43 495 736 719 |
| 0.95 | 1.82 591 977 025 | 0.49 291 086 358 | -0.38 127 732 889 | 1.91 942 912 742 | 0.58 057 816 906 | -0.40 982 961 702 |
| 1 | 1.85 009 851 224 | 0.47 443 018 048 | -0.35 830 925 451 | 1.94 795 551 523 | 0.56 066 834 203 | -0.38 690 574 350 |

## Tables of  H(u) , H$^{/}$(u)  and H$^{//}$(u)
### (determined
### from
### equations(15a,74 & 75))
### (table -7)

|  | c = 0.95 | | | c = 0.975 | | |
|---|---|---|---|---|---|---|
| u | H(u) | H$^{/}$(u) | H$^{//}$(u) | H(u) | H$^{/}$(u) | H$^{//}$(u) |
| 0.0 | 1.00 000 000 000 | +∞ | -∞ | 1.00 000 000 000 | +∞ | -∞ |
| 0.05 | 1.11 150  780 933 | 1.81 961 340 886 | -7.65 311 383 064 | 1.11 941 844 758 | 1.97 634 127 388 | -7.53 874 376 961 |
| 0.1 | 1.19 523 180 435 | 1.56 101 570 044 | -3.65 010 974 638 | 1. 21 114 615 023 | 1. 72 466 662 292 | -3.50 404 263 808 |
| 0.15 | 1.26 935 437 885 | 1.41 406 832 050 | -2.40 591 043 505 | 1.29 363 087 512 | 1.58 481 586 592 | -2.26 999 464 194 |
| 0.2 | 1.33 733 676 938 | 1.31 013 190 209 | -1.81 189 715 910 | 1.37 031 430 638 | 1.48 729 809 522 | -1.69 126 413 083 |
| 0.25 | 1.40 073 988 566 | 1.22 885 959 491 | -1.46 597 052 048 | 1.44 272 008 013 | 1.41 167 239 720 | -1.36 059 954 224 |
| 0.3 | 1.46 045 389 988 | 1.16 158 326 276 | -1.23 927 492 123 | 1.51 170 041 612 | 1.34 930 531 123 | -1.14 806 388 507 |
| 0.35 | 1 .51 705 625 064 | 1 .10 384 724 968 | -1.07 849 034 201 | 1 .57 779 738 922 | 1 .29 580 294 111 | -1.00 013 862 868 |
| 0.4 | 1.57 095 413 754 | 1.05 307 216 632 | -0.95 779 498 268 | 1.64 138 604 153 | 1.24 865 047 401 | -0.89 103 909 042 |
| 0.45 | 1.62 245 214 427 | 1.00 763 483 039 | -0.86 325 643 628 | 1.70 274 194 563 | 1.20 628 548 697 | -0.80 693 409 063 |
| 0.5 | 1.67 178 828 697 | 0.96 644 787 906 | -0.78 673 463 991 | 1.76 207 701 008 | 1.16 767 590 502 | -0.73 979 662 865 |
| 0.55 | 1.71 915 491 079 | 0.92 874 631 959 | -0.72 317 217 769 | 1.81 956 009 661 | 1.13 270 645 547 | -0.68 467 747 898 |
| 0.6 | 1.76 471 160 260 | 0.89 397 025 320 | -0.66 926 830 610 | 1.87 532 966 333 | 1.09 906 178 690 | -0.63 837 428 708 |
| 0.65 | 1.80 859 358 555 | 0.86 169 626 096 | -0.62 277 969 805 | 1.92 950 191 338 | 1.06 815 837 399 | -0.59 873 150 665 |
| 0.7 | 1.85 091 740 749 | 0.83 159 522 423 | -0.58 212 881 471 | 1.98 217 626 386 | 1.03 910 285 659 | -0.56 424 937 853 |
| 0.75 | 1.89 178 493 218 | 0.80 340 528 196 | -0.54 617 341 827 | 2.03 343 914 233 | 1.01 166 547 200 | -0.53 385 447 209 |
| 0.8 | 1.93 128 622 394 | 0.77 691 387 371 | -0.51 406 526 761 | 2.08 336 669 918 | 0.98 566 251 930 | -0.50 675 946 741 |
| 0.85 | 1. 96 950 168 671 | 0.75 194 544 963 | -0.48 516 025 639 | 2.13 202 679 426 | 0.96094 443 818 | -0.48 237 429 474 |
| 0.9 | 2.00 650 368 654 | 0.72 835 283 457 | -0.45 895 956 433 | 2.17 948 048 402 | 0.93 738 749 459 | -0.46 024 817 232 |
| 0.95 | 2.04 235 780 680 | 0.70 601 101 721 | -0.43 507 004 486 | 2.22 578 315 647 | 0.91 488 785 014 | -0.44 003 076 052 |
| 1 | 2.07 712 383 665 | 0.68 481 258 977 | -0.41 317 692 562 | 2.27 098 541 258 | 0.89 335 724 650 | -0.42 144 552 017 |

**Tables of  H(u) , H$'$(u)  and H$''$(u)**
**(determined**
**from**
**equations (16a,79& 80)**
**(table -8)**

c=1.00

| u | H(u) | H$'$(u) | H$''$(u) |
|---|---|---|---|
| 0.0 | 1.00 000 000 000 | +∞ | -∞ |
| 0.05 | 1.13 657 484 173 | 2.33 706 572 499 | -6.72 070 082 043 |
| 0.1 | 1.24 735 043 713 | 2.12 552 340 854 | -2.72 359 699 103 |
| 0.15 | 1.35 083 358 726 | 2.02 350 707 216 | -1.53 614 738 251 |
| 0.2 | 1.45 035 140 709 | 1.96 164 572 714 | -0.99 784 464 484 |
| 0.25 | 1.54 732 622 812 | 1.91 979 438 791 | -0.70 227 490 006 |
| 0.3 | 1.64 252 225 848 | 1.88 955 236 350 | -0.52 080 318 072 |
| 0.35 | 1.73 640 371 932 | 1.86 670 241 055 | -0.40 084 735 042 |
| 0.4 | 1.82 927 559 701 | 1.84 886 745 874 | -0.31 726 244 572 |
| 0.45 | 1.92 134 958 544 | 1.83 459 604 282 | -0.25 666 274 753 |
| 0.5 | 2.01 277 876 364 | 1.82 294 812 769 | -0.21 134 149 768 |
| 0.55 | 2.10 367 740 351 | 1.81 328 667 785 | -0.17 658 736 828 |
| 0.6 | 2.19 413 301 282 | 1.80 516 415 003 | - 0.14 938 024 508 |
| 0.65 | 2.28 421 402 476 | 1.79 825 674 452 | -0.12 770 805 911 |
| 0.7 | 2.37 397 490 591 | 1.79 232 438 331 | -0.11 018 679 242 |
| 0.75 | 2.46 345 966 186 | 1.78 718 533 331 | -0.09 583 837 973 |
| 0.8 | 2.55 270 431 011 | 1.78 269 955 514 | -0.08 395 555 017 |
| 0.85 | 2.64 173 866 589 | 1.77 875 745 457 | -0.07 401 650 233 |
| 0.9 | 2.73 058 765 805 | 1.77 527 209 016 | -0.06 562 948 323 |
| 0.95 | 2.81 927 231 610 | 1.77 217 365 584 | -0.05 849 579 987 |
| 1 | 2.90 781 052 218 | 1.76 940 549 822 | -0.05 238 456 191 |

We have compared our result  of H- functions, its first derivative and second derivative with the results so far available only for the particular cases for u=0.05.0.5 and1.0  For c=0.2,0.4,0.6,0.8,0.9 and 1.0. in tables no- 10 and 11.

### ( comparison table –9)

| | | Comparison | of available | results | with results | in    this | paper |
|---|---|---|---|---|---|---|---|
| | | c =0.2 | c= 0.4 | c=0.6 | c=0.8 | c=0.9 | c=1.00 |
| | u=0.05 | | | | | | |
| Chandrasekhar | H(u) | 1.01 608 | 1.03 422 | 1.05 544 | 1.08 200 | 1.09 99 | 1.13 68 |
| 1950,page 125 | H′(u) | | | | | | |
| [1] | H″(u) | | | | | | |
| | | | | | | | |
| Stibbs and Weir | H(u) | 1.01 605 3 | 1.03 416 3 | 1.05 531 7 | 1.08 191 4 | 1.09 967 8 | 1.13 657 5 |
| 1957,page 518 | H′(u) | | | | | | |
| [12] | H″(u) | | | | | | |
| | | | | | | | |
| Hovenier,Mee,He er | H(u) | 1.01 605 | 1.03 416 | 1.05 532 | 1.08 191 | 1.09 968 | 1.13 657 |
| 1988,page 196 | H′(u) | 0.22 587 | 0.49 368 | 0.82 506 | 1.27 296 | 1.59 419 | 2.33 707 |
| [31] | H″(u) | -1.82 894 | -3.66 843 | -5.46 214 | -7.04 906 | -7.58 719 | -6.72 070 |
| | | | | | | | |
| Das and Bera From eqs | H(u) | 1.01 605 299 583 | 1.03 416 322 685 | 1.05 531 672 613 | 1.08 191 451 387 | 1.09 967 828 295 | 1.13 657 484 173 |
| 15a,74 & 75 | H′(u) | 0.22 586 684 758 | 0.49 368 012 768 | 0.82 506 115 268 | 1.27 296 409 722 | 1.59 418 923 918 | 2.33 706 572 499 |
| In this paper | H″(u) | -1.82 893 782 344 | -3.66 842 935 352 | -5.46 213 733 606 | -7.04 906 473 873 | -7.58 718 876 030 | -6.72 070 082 043 |

**(comparison table –10)**

| | | Comparison | of available | results | with results | in this paper | |
|---|---|---|---|---|---|---|---|
| | | c =0.2 | c= 0.4 | c=0.6 | c=0.8 | c=0.9 | c=1.00 |
| | u=0.5 | | | | | | |
| Chandrasekhar | H(u) | 1.06 117 | 1.13 918 | 1.24 581 | 1.41 32 | 1.55 60 | 2.01 28 |
| [1],page 125 | H′(u) | | | | | | |
| | H″(u) | | | | | | |
| Stibbs and Weir | H(u) | 1.06 117 6 | 1.13 919 2 | 1.24 5807 | 1.41 326 2 | 1.55 603 3 | 2.01 277 8 |
| [12],page 518 | H′(u) | | | | | | |
| | H″(u) | | | | | | |
| Hovenier,Mee,H eer | H(u) | 1.06 118 | 1.13 919 | 1.24 581 | 1.41 326 | 1.55 603 | 2.01 278 |
| [31],page 196 | H′(u) | 0.05 154 | 0.12 737 | 0.24 973 | 0.48 699 | 0.73 435 | 1.82 295 |
| | H″(u) | -0.10 062 | -0.23 120 | -0.40 550 | -0.063 980 | -0.76 441 | -0.21 134 |
| Das and Bera from eqs (15a) , 74 & 75 | H(u) | 1.06 117 662 957 | 1.13 919 205 779 | 1.24 580 715 523 | 1.41 326 256 478 | 1.55 603 379 696 | 2.01 277 876 364 |
| | H′(u) | 0.05 154 034 237 | 0.12 736 803 214 | 0.24 973 013 986 | 0.48 699 231 899 | 0.73 434 958 190 | 1.82 294 812 769 |
| In this paper | H″(u) | -0.10 062 193 515 | -0.23 120 387 370 | -0.40 550 469 570 | -0.63 980 466 634 | -0.76 440 881 931 | -0.21 134 149 768 |

**( comparison table –11)**

| | | Comparison | of available | results | with results | in this paper | |
|---|---|---|---|---|---|---|---|
| | u=1.0 | c =0.2 | c= 0.4 | c=0.6 | c=0.8 | c=0.9 | c=1.00 |
| Chandrasekhar | $H(u)$ | 1.07 864 | 1.18 337 | 1.33 541 | 1.59 82 | 1.85 01 | 2.90 78 |
| [1],page 125 | $H'(u)$ | | | | | | |
| | $H''(u)$ | | | | | | |
| Stibbs and Weir | $H(u)$ | 1.07 864 4 | 1.18 338 0 | 1.33 540 6 | 1.59 821 9 | 1.85 009 8 | 2.90 780 9 |
| [12],page 518 | $H'(u)$ | | | | | | |
| | $H''(u)$ | | | | | | |
| Hovenier,Mee,He | | | | | | | |
| er | $H(u)$ | 1.07 865 | 1.18 338 | 1.33 541 | 1.59 822 | 1.85 010 | 2.90 781 |
| [31],page 196 | $H'(u)$ | 0.02 387 | 0.06 180 | 0.12 958 | 0..28 307 | 0.47 443 | 1.76 941 |
| | $H''(u)$ | -0.02 977 | -0.07 313 | -0.14 105 | - 0.26 05 7 | -0.35 831 | -0.05 238 |
| Das and Bera From eqs | $H(u)$ | 1.07 864 506 335 | 1.18 338 066 380 | 1.33 540 666 947 | 1.59 821 951 461 | 1.85 009 851 224 | 2.90 781 052 218 |
| (15a,74 &75) | $H'(u)$ | 0.02 387 090 108 | 0.06 180 366 962 | 0.12 957 561 195 | 0.28 307 108 419 | 0.47 443 018 048 | 1.76 940 549 822 |
| In this paper | $H''(u)$ | -0.02 977 000 787 | -0.07 312 566 678 | -0.14 104 697 297 | -0.26 057 177 876 | -0.35 830 925 451 | -0.05 238 456 191 |

We have compared the results of moments determined from this new H-function with the results of others as follows in tables no −12,13,14 and 15.

### ( comparison table-12)

| | Moment | Table | of | H-function |
|---|---|---|---|---|
| | c | 0.1 | 0.2 | 0.3 |
| moments obtained | zeroth moment | 1.02 633 40 | 1.05 572 81 | 1.08 893 31 |
| from | first moment | 0.51 560 9 | 0.53 315 4 | 0.55 312 3 |
| (e.g Chandrasekhar[1] | 2ndmoment | 0.34 435 7 | 0.35 678 7 | 0.37 098 5 |
| ,page126,328) | 2nd mom/1st mom | | | |
| moments obtained | zeroth moment | 1.02 63 | 1.05 57 | 1.08 89 |
| from | first moment | 0.51 56 | 0.53 32 | 0.55 31 |
| (e.g Ivanov | 2ndmoment | 0.34 44 | 0.3568 | 0.37 10 |
| [25],page 130) | 2nd mom/1st mom | | | |
| moments obtained | zeroth moment | | 1.05 573 | |
| from | first moment | | 0.53 315 | |
| (e.g Hovenier | 2ndmoment | | 0.35 679 | |
| [31],page 196) | 2nd mom/1st mom | | | |
| moments determined | zeroth moment | 1.02 633 403 899 | 1.05 572 809 000 | 1.08 893 315 644 |
| from equation(44 ) | first moment | 0.51 561 065 494 | 0.53 315 418 225 | 0.55 312 108 172 |
| by Das and Bera | 2ndmoment | 0.34 435 834 332 | 0.35 678 778 614 | 0.37 098 417 291 |
| In this paper | 2nd mom/1st mom | 0.66 786 506 451 | 0.66 920 188 946 | 0.67 071 060 059 |

**( comparison table-13)**

| | Moment | Table | of | H-function |
|---|---|---|---|---|
| | c | 0.4 | 0.5 | 0.6 |
| moments obtained | zeroth moment | 1.12 701 67 | 1.17 157 29 | 1.22 514 82 |
| from | first moment | 0.57 621 0 | 0.60 349 5 | 0.63 663 6 |
| (e.g Chandrasekhar[1] | 2ndmoment | 0.38 746 6 | 0.40 703 0 | 0.43 092 2 |
| page126,328) | 2nd mom/1st mom | | | |
| moments obtained | zeroth moment | 1.12 70 | 1.17 16 | 1.22 51 |
| from | first moment | 0.57 62 | 0.60 35 | 0.63 66 |
| (e.g Ivanov | 2ndmoment | 0.38 75 | 0.40 70 | 0.43 09 |
| [25],page 130 ) | 2nd mom/1st mom | | | |
| moments obtained | zeroth moment | 1.12 702 | | 1.22 515 |
| from | first moment | 0.57 621 | | 0.63 663 |
| (e.g Hovenier | 2ndmoment | 0.38 747 | | 0.43 092 |
| [31],page 196) | 2nd mom/1st mom | | | |
| moments determined | zeroth moment | 1.12 701 665 379 | 1.17 157 287 525 | 1.22 514 822 655 |
| from equation(44 ) | first moment | 0.57 621 284 964 | 0.60 348 426 311 | 0.63 663 257 762 |
| by Das and Bera | 2ndmoment | 0.38 746 773 474 | 0.40 702 364 786 | 0.43 092 070 033 |
| In this paper | 2nd mom/1st mom | 0.67 243 855 284 | 0.67 445 610 886 | 0.67 687 503 825 |

**( comparison table-14)**

| | Moment<br>c | Table<br>0.7 | of<br>0.8 | H-function<br>0.9 |
|---|---|---|---|---|
| moments obtained from (e.g Chandrasekhar [1],page126,328) | zeroth moment<br>first moment<br>2ndmoment<br>2nd mom/1st mom | 1.29 222 13<br>0.67 867 4<br>0.46 142 3 | 1.38 196 6<br>0.73 580 8<br>0.50 321 8 | 1.51 949 4<br>0.82 531 8<br>0.56 944 9 |
| moments obtained from (e.g Ivanov [25],page 130) [ | zeroth moment<br>first moment<br>2ndmoment<br>2nd mom/1st mom | 1.29 22<br>0.67 87<br>0.4614 | 1.38 20<br>0.73 58<br>0.50 32 | 1.51 95<br>0.82 53<br>0.56 94 |
| moments obtained from (e.g Hovenier [31],page 196) | zeroth moment<br>first moment<br>2ndmoment<br>2nd mom/1st mom | | 1.38 197<br>0.73 582<br>0.50 322 | 1.51 949<br>0.82 532<br>0.56 945 |
| moments determined from equation(44  ) by Das and Bera In this paper | zeroth moment<br>first moment<br>2ndmoment<br>2nd mom/1st mom | 1.29 222 126 427<br>0.67 866 782 287<br>0.16 141 989 137<br>0.67 989 062 664 | 1.38 196 601 125<br>0.73 581 523 568<br>0.50 322 379 413<br>0.68 389 966 629 | 1.51 949 385 330<br>0.82 531 575 014<br>0.56 944 861 767<br>0.68 997 667 568 |

## ( comparison  table-15)

| | Moment | Table | of | H-function | |
|---|---|---|---|---|---|
| | c | 0.925 | 0.95 | 0.975 | 1 |
| moments obtained | zeroth moment | 1.57 003 0 | 1.63 451 2 | 1.72 694 6 | 2.00 000 0 |
| from | first moment | 0.85 873 4 | 0.90 186 4 | 0.96 447 1 | 1.15 470 1 |
| (e.g Chandrasekhar | 2ndmoment | 0.59 440 4 | 0.62 678 5 | 0.67 413 4 | 0.82 035 2 |
| [1],page126,328) | 2nd mom/1st mom | | | | |
| moments obtained | zeroth moment | 1.57 00 | 1.63 45 | 1.72 69 | 2.00 00 |
| from | first moment | 0.85 88 | 0.90 19 | 0.96 45 | 1.15 47 |
| (e.g Ivanov | 2ndmoment | 0.59 44 | 0.62 68 | 0.67 41 | 0.82 04 |
| 25,page130) | 2nd mom/1st mom | | | | |
| moments obtained | zeroth moment | | 1.63 451 | | 2.00 000 |
| from | first moment | | 0.90 188 | | 1.15 470 |
| (e.g Hovenier | 2ndmoment | | 0.62679 | | 0.82 035 |
| [31],page196) | 2nd mom/1st mom | | | | |
| moments determined | zeroth moment | 1.57 002 966 756 | 1.63 451 200 474 | 1.72 694 588 101 | 2.00 000 000 000 |
| from equation( 44 & 58)) | first moment | 0.85 876 731 020 | 0.90 187 602 626 | 0.96 448 124 829 | 1.15 470 053 838 |
| by Das and Bera | 2ndmoment | 0.59 442 789 984 | 0.62 679 443 579 | 0.67 414 203 488 | 0.82 035 247 952 |
| In this paper. | 2nd mom/1st mom | 0.69 218 738 625 | 0.69 498 957 455 | 0.69 896 852 435 | 0.71 044 608 732 |

## Conclusion :

We believe that the method of numerical solution of H-functions , and its derivatives from an exclusively  new form presented here is sufficiently different from others in the literature of H – function in radiative transfer   to warrant its communication . Consequently it is hoped that this paper will meet the commitment made by one of the authors in a paper and fill up the gap still available for enriching   the theory of  radiative transfer on H-function  .The scope of this paper has been limited to a discussion of the H-functions which arise in the case of the coherent isotropic  scattering of radiation in a semi infinite atmosphere and to the presentation of the results of calculation in a form suited to the need for rapid computation of the H-function , its first derivative and second derivative , pole  and their moments for any value of particle albedo of scattering for different direction parameter . The method described in this paper , however be extended with further elaboration to the scattering which arise in the theory of the formation of multiplet lines by non coherent isotropic and anisotropic scattering .

**Acknowledgement** :

We express our sincere thanks to the Librarian and support officials of Saha Institute of Nuclear Physics , Salt Lake, Kolkata , India and of Satyen Bose Institute of Basic Sciences , Salt Lake, Kolkata , India and Department of Mathematics , Heritage Institute of Technology , Anadapur, West Bengal , India for their whole hearted support .We sincerely thanks to Sri Soumyava Das, a B.Tech Student of Computer Science of Jadavpur University , West Bengal , for his computer programming to arrive at this accuracy .

.

**The End.**

.

.

.